\def\odd{$O(d,d)$}
\def\oddh{$O(d,d+16)$}
\def\be{\begin{equation}}
\def\ee{\end{equation}}
\def\beq{\begin{eqnarray}}
\def\eeq{\end{eqnarray}}
\def\wt{\widetilde}
\def\del{\partial}
\def\ds{\displaystyle}
\documentstyle[12pt]{article}
\advance\textheight by \topskip
\begin{document}
\renewcommand{\thefootnote}{\fnsymbol{footnote}}
\vspace{-0.2cm}
\begin{flushright}
TIFR/TH/94-26\\
hep-th/9408060
\end{flushright}
\vspace{-0.5cm}
\begin{center}
{\large\bf O(d,d;R) Deformations of Complex Structures  \\ and \\
Extended Worldsheet Supersymmetry}\\
\vskip 2 cm
{\bf S. F. Hassan}\footnote{ E-mail address:
{\tt fawad@surya1.cern.ch}}\footnote{Address after November 1994:
Theory Division, CERN, CH-1211, Geneva-23, Switzerland.}
\vskip 0.05cm
{Tata Institute of Fundamental Research\\ Homi Bhabha Road\\
Bombay 400 005, India}\\
\end{center}
\vskip 1.0 cm
\centerline{\bf ABSTRACT}
\begin{quotation}
\noindent
It is shown that the $O(d,d;{\rm R})$ deformations of the superstring
vacua and the $O(d,d+16;{\rm R})$ deformations of the heterotic string
vacua preserve extended worldsheet supersymmetry and, hence, generate
superconformal deformations.  The transformations of the complex
structures are given explicitly and the action of the discrete duality
subgroup is discussed. The results are valid when the complex
structures are independent of the $d$ coordinates which appear in the
transformations.  It is shown that generic deformations do not preserve
the known superfield formulations of $(2,2)$ extended supersymmetry.
The analysis is performed by decomposing the transformations in terms
of the metric vielbein and by introducing space-time connections
induced due to the non-linear action of the $O(d,d;{\rm R})$ and
$O(d,d+16;{\rm R})$ deformations on the background fields.
\end{quotation}

\newpage
\renewcommand{\thefootnote}{\arabic{footnote}}
\setcounter{footnote}{0}

\section{Introduction}
String theories with extended supersymmetry on the worldsheet are
important because of the restrictions which the extra global
supersymmetries impose on the target space geometry
\cite{ZAF,GHR,HW,HP,dN} as well as their
natural relation to space-time supersymmetry \cite{SenN=2}. The study
of N=2 theories, and the profound structures associated with them has
been an area of intense investigation in recent years. On the other
hand, it is well known that the string vacua, considered as conformal
field theories, in general, have deformations under which the geometry
of the target space changes. An important case is when the background
fields have translational invariance along $d$ of the space
coordinates.  The string vacua, then, admit deformations generated by
the group $O(d,d;{\rm R})$ \cite{MV,Sen,GR,Def},
which, in heterotic string theory with abelian gauge fields is
generalized to $O(d,d+16;{\rm R})$ \cite{twist}\cite{JMJS}. This is a
generalization of Narain's group to space-time dependent background
fields \cite{Narain}\cite{GRVM}.  These deformations include the
discrete duality transformations which are symmetries of the
underlying conformal field theory \cite{Busc}\cite{RV}\cite{GR}. For
an exhaustive set of references on this subject see \cite{Review}.
Now, consider a non-linear $\sigma$-model with extended supersymmetry
on the worldsheet which also admits \odd\, (or \oddh\,) deformations.
These deformations change the background fields in a highly
non-trivial manner and, therefore, interfere with the constraints
which the extended worldsheet supersymmetry imposes on the target
space geometry.  The issue of interest is to study the effect of the
deformations of the background fields on the extended worldsheet
supersymmetry.

The $N=1$ local supersymmetry on the worldsheet does not impose any
restrictions on the background fields and is, therefore, trivially
preserved under the above mentioned deformations. The extended
supersymmetries, however, require the existence of complex structures
on the target manifold thereby restricting its geometry. If the theory also
has some isometries, then it admits deformations which change the
target space geometry and field configuration, while, preserving the
worldsheet conformal invariance. It has been known for sometime that
theories with extended $N=2$ supersymmetry which can be represented in
terms of chiral and twisted chiral superfields and which also have an
isometry, admit a discrete duality transformation \cite{GHR}. This
duality transformation converts a chiral superfield into a twisted
chiral one (or {\it vice versa}) and also changes the target space
geometry. It has been shown that this duality is the same as the usual
duality transformation which is a discrete subgroup of the \odd\,
group when the latter is applied to $N=2$ theories which admit a manifest
superfield representation\cite{RV}\cite{KKL}. A generic class of such
theories in four dimensions were considered in \cite{KKL} and the
duality transformation was combined with gauge transformations of the
anti-symmetric tensor field to produce a family of non-trivial \odd\
deforamations of the background fields. Since at any step the
transformation is compatible with supersymmetry, it is insured that
the one-parameter family of theories thus obtained also has an
extended supersymmetry. However, these theories may no longer have a
representation in terms of chiral and twisted chiral
superfields.  In \cite{KR}, duality with respect to one isometry
direction was considered in the more general case, when a superfield
representation is not necessarily known, and the dual complex
structures were obtained. Strictly speaking, these results are valid
only when the complex structures do not depend on the coordinate with
respect to which duality is performed \footnote{See the note added at
  the end regarding the recent results when this is not the case.}.
Since non-trivial \odd\ transformations can be obtained by
intertwining duality transformations with general coordinate
transformations and gauge transformations of the anti-symmetric tensor
field, it is expected that they should also preserve the extended
worldsheet supersymmetry. However, the action of a generic,
non-trivial \odd\ or \oddh\ transformation on the complex
structures and the associated supersymmetry is not clear.  In this
paper, we undertake a study of this problem by investigating the
effect of deformations (including duality transformations) of the
superstring and heterotic string backgrounds on the constraints that
the supersymmetries impose on the target  manifold. However, we
restrict ourselves to the case when the complex
structures are independent of the $d$ coordinates with respect to
which the deformations are performed. We obtain the transformation
properties of the complex structures and show that the deformations
preserve extended supersymmetries on the worldsheet.  Hence, they
correspond to marginal deformations of the underlying superconformal
field theory. They, however, do not preserve the K\"{a}hler structure
\cite{ZAF} and the product structure \cite{GHR}\cite{HP} on the target
manifold. The explicit form of the transformations we have used are
accurate to one-loop level in the $\sigma$-model perturbation theory,
although corrections are known to exist to all orders
\cite{Sen}\cite{twist}. The calculation is simplified by introducing
some quantities which transform nicely under \odd\, and \oddh\,
deformations.

The paper is organized as follows: In section $2$, we linearize the
action of the $O(d,d)$ transformations using the target space
vielbeins. We, then, construct two connections induced by these
transformations which are related by the worldsheet parity
transformation. In section $3$, we consider non-linear $\sigma$ models
with extended $N=2$ and $N=4$ supersymmetries and translational
invariance along $d$ of the target space coordinates. We find the
transformations of the complex structures under \odd\, deformations
and using the connections introduced in section $2$, show that the
deformations preserve the extended worldsheet supersymmetry.  In
section $4$, we consider, in more detail, the transformations of the
complex structures in relation to manifest superfield representation
of theories with extended $N=2$ supersymmetry. It is shown that,
generically, the deformed theories do not have a manifest superfield
representation. The case of discrete duality transformation with
respect to $d$ isometries is discussed with reference to its relation
to a duality known in the $N=2$ theories. In section 5, we linearize
the action of the \oddh\, group on the heterotic string backgrounds
and, using the induced connection, show the invariance of the extended
$(0,2)$ and $(0,4)$ supersymmetries under the deformations. The
results are summarized in section $6$.

\section{Linearization of the $O(d,d)$ Transformations and
the Induced Connections}

In this section we will rewrite the action of the \odd\, group, which
deforms string vacua with $d$ isometries, in terms of the target space
vielbeins and review its relevant aspects for later reference. Then,
we introduce two matrices, $Q_\pm$, which implement the deformations on
the background fields and construct two \odd\, induced connections.
These will be used in our analysis of extended worldsheet
supersymmetry in section $3$.

It is known that when the background fields, $G_{MN}(X)$\,,
$B_{MN}(X)$\, and $\Phi(X)$ are independent of some $d$ (out of
$D$\,) of the space-time coordinates, then the low-energy effective
action is invariant under a set of non-trivial transformations which
are generated by the non-linear action of an \odd\, group on the
background fields \cite{MV}\cite{Sen}\cite{GR}.  In the case of
heterotic string theory, this deformation group is enlarged to
$O(d,d+p)$ for a configuration of the gauge fields which commutes with
$p$ of the Cartan generators of the gauge group
\cite{twist}\cite{JMJS}.  These transformations, on one hand, change
the geometry of space-time and, on the other, generate deformations in
the underlying conformal field theory \cite{Def}. The corresponding
conformal field theory moduli can be identified with quantities like
the axion charge, angular momentum and the electric charge in the
associated space-time theory. In this section we will discuss the
\odd\, transformations in superstring theory and will return to the
case of heterotic string theory in section $5$. The conventions
adopted are as follows: The $D$-dimensional space-time indices are
written as $K,L,M,\ldots$ and the corresponding tangent space indices
are written as $A,B,C,\ldots$.  All fields are assumed to be
independent of $d$ of the space coordinates with indices denoted by
$l,m,n,\ldots$. The remaining $D-d$ coordinates, on which the
background fields can depend, include the time direction and carry
indices $\mu,\nu,\gamma,\ldots$. We will, however, use the matrix
notation and suppress the indices whenever possible. In that case, the
first $d$ rows and $d$ columns are labeled by indices $l,m,n,\ldots$
and the remaining ones are labeled by the indices
$\mu,\nu,\gamma,\ldots$.

The object which transforms nicely under the \odd\, transformations is
a $2D$ dimensional matrix, $M$, constructed out of the background
fields $G_{MN}$ and $B_{MN}$ as
\be
\label{M}
M = \left(\begin{array}{cc}
G^{-1} & 1_D - G^{-1} B \\ 1_D + B G^{-1} & G - B G^{-1} B
\end{array}\right)
\ee
To obtain an action on $M$, the group \odd\, is considered as
a subgroup of the larger group $O(D,D)$ in its fundamental
representation. If $\Omega$ is an element of the \odd\, group
constructed in this way, then the matrix $M$ transforms under the
adjoint action of $\Omega$,
\be
\label{oddM}
\wt M =\,\, \Omega M \Omega^T, \qquad \Omega \in O(d,d) \subset O(D,D)
\ee
Since the fields $G_{MN}$ and $B_{MN}$ are uniquely determined by $M$,
their transformation can be obtained from that of $M$. At the one-loop
level in the $\sigma$-model perturbation theory, this is also
supplemented by an appropriate transformation of the dilaton field
$\Phi$ which is not relevant for our purposes.  The representation is
chosen such that the equation defining the group elements $\Omega$
takes the form
\be
\label{L}
\Omega^T\,L\,\Omega = L\,,\quad
L=\left(\begin{array}{cc}0&1_D\\1_D&0\end{array}\right)
\ee
This shows that for the purpose of extracting the transformations of
the background fields, $M$ is arbitrary up to the addition of a
multiple of $L$.

The embedding of $\Omega$ in $O(D,D)$ is given by the following
parametrization
\be
\Omega =
\left(\begin{array}{cc}A_1 & C_1 \\C_2 & A_2
\end{array}\right);\,\,
A_i =\left(\begin{array}{ll}{\cal A}_i & 0\\0 &
1_{D-d}\end{array}\right),\,
C_i =\left(\begin{array}{ll}{\cal C}_i & 0\\ 0 & 0_{D-d}\end{array}
\right),
\ee
where,
$$
\left(\begin{array}{cc}{\cal A}_1 & {\cal C}_1 \\ {\cal C}_2 &
{\cal A}_2\end{array} \right) \in O(d,d)
$$
A general \odd\, transformation can be parametrized in terms of its
action on the background fields. The group elements given by ${\cal
  A}_2=({\cal A}_1^T)^{-1},{\cal C}_1={\cal C}_2=0$\,, correspond to
$GL(d,R)$ transformations and the ones given by ${\cal A}_2={\cal
  A}_1=1,{\cal C}_1=0,{\cal C}_2=-{\cal C}_2^T$\,, correspond to
constant gauge transformations of the antisymmetric tensor field,
$B_{MN}$. The non-trivial deformations of the backgrounds are
generated by elements from $O(d)\times O(d)/O(d)$.  The $O(d)\times
O(d)$ subgroup is parametrized as
\be
\label{SR}
\Omega =\frac{1}{2}\left(\begin{array}{cc} S+R & S-R \\ S-R & S+R
\end{array}\right)\,,\, {\rm where},\,
S(R)=\left(\begin{array}{cl} {\cal S}({\cal R}) & 0 \\
0 & 1_{D-d}
\end{array}\right), \,\, {\cal S}, {\cal R} \in O(d).
\ee
The diagonal $O(d)$ subgroup, given by ${\cal S}={\cal R}$, generates
ordinary rotations and is already included in $GL(d,R)$. To get
non-trivial deformations, one considers (\ref{SR}) modulo this
subgroup. The usual duality transformations \cite{Busc} with respect
to isometries along coordinates $X^i$, are contained in a discrete
subgroup generated by
\be
\label{dual}
{\cal S}=1\,,\, {\cal R} = 1-2\varepsilon_i\,,\quad {\rm where},\,
(\varepsilon_i)_{jk}=\delta_{ij}\delta_{ik}
\ee
The full generalized duality group, $O(d,d;{\rm Z})$, is generated by
(\ref{dual}), along with $GL(d,{\rm Z})$ and discrete gauge
transformations of $B_{MN}$ \cite{GR}. In the following, however, we
will only refer to (\ref{dual}) as duality transformations.

To simplify the manipulations, in the following, we rewrite the above
transformations in terms of the metric vielbein.  Using the inverse
vielbein, $e$, and a matrix $K$ defined as
\be
 G^{-1}= e\, \eta\, e^T ,\qquad K = G + B ,
\ee
we construct a $2D\times D$ rectangular matrix $\xi$ as
\be
\xi = \left(\begin{array}{c} e \\ Ke \end{array}\right)
\ee
This is related to $M$ by
$$
M = \xi\, \eta\, \xi^T =
\left(\begin{array}{c} e \\ K e \end{array}\right)\!\!
\begin{array}{c}\eta\\ \, \end{array}\!\!
\begin{array}{c}
\left(\begin{array}{cc} e^T &  e^T K^T \end{array}\right)\\ \,
\end{array}
$$
and transforms, under \odd\,, as
\be
\label{odd}
\wt\xi= \Omega\,\xi\, , \qquad \Omega \in O(d,d) \subset O(D,D)
\ee
Note that the Lorentz indices are not affected by the
transformations\footnote{When $S=R$, one may perform compensatory
transformations on the Lorentz indices so that the flat vielbein is
invariant under rotations.}. In the remaining part of this section, we
will describe some constructions which are used for our analysis of
extended worldsheet supersymmetry in section $3$.

Since the extended worldsheet supersymmetries are manifestly invariant
under $GL(d,R)$ and the $B_{MN}$ gauge transformations, in the
following, we will mainly concentrate on the non-trivial deformations
generated by the $O(d)\times O(d)$\, subgroup.  As argued in the
previous section, it is expected that these deformations also preserve
the extended supersymmetry, though their effect on the complex
structures and, therefore, on the supersymmetry charges is far from
clear. Under the action of this subgroup, which is parametrized by the
matrices $S$ and $R$, the vielbein $e$, the metric G, and the matrix
$K=G+B$ transform as
\be
\label{eGK}
\begin{array}{ccl}
\wt e &=& Q_-(S,R)\, e \\
\wt G^{-1} &=& Q_-(S,R)\, G^{-1} Q_-^T(S,R)  \\
\wt K &=& Q_-(S,-R)\,Q_-^{-1}(S,R)
\end{array}
\ee
where,
\be
\label{Qm}
\begin{array}{l}
Q_-(S,R)={\ds\frac{1}{2}}\bigg(\, S+R+(S-R)\, K\, \bigg) \\ \\
Q_-^{-1}(S,R)={\ds\frac{1}{2}}\bigg(\, S^T+R^T+(S^T-R^T)\,{\wt K} \bigg)
\end{array}
\ee
The second equation above follows from the fact that
$\Omega^{-1}(S,R)=\Omega(S^T,R^T)$ and the subscript for $Q$ has been
chosen in anticipation of the transformation properties of the complex
structures to be discussed in the next section. Note that though
\odd\, is a global transformation, the matrix $Q_-$ which implements
the transformation becomes local through its
dependence on $K(X)= G(X) + B(X)$. Also, since the conditions for the
existence of extended supersymmetry on the worldsheet involve
space-time derivatives of the background fields, it is desirable to
find convenient expressions for the transformation of such
quantities. To this end, first note that under the transformation
(\ref{odd}), the quantity \,$\xi^T L\,\del_\mu \xi$\, is
invariant. Now, consider a vector field $V(X)$ which under \odd\,
transforms as $\wt V = Q_- V$. Corresponding to this transformation,
we can construct an \odd\, induced connection \footnote{As expected,
this connection has zero curvature.} as $e\,\del_\mu e^{-1}$.  However,
it is more useful to add an \odd\-covariant piece to this expression
and construct a new induced connection, $\omega^-_\mu$, given by
\be
\label{connecm}
\omega^-_\mu = \frac{1}{2}\, e\eta\xi^T L \del_\mu\xi e^{-1}
- e\del_\mu e^{-1} ,
\ee
in terms of which,
\be
\del_\mu K = 2\, G \, \omega^-_\mu
\ee
Under an \odd\, deformation, this connection transforms to
\be
\label{trconnecm}
\wt \omega^-_\mu =  Q_-\, \omega^-_\mu Q_-^{-1}- \del_\mu Q_- Q_-^{-1}
\ee
It should be emphasized that $\omega^-_\mu$ is a connection in the
space-time sense but is not the same as the natural torsion-full
connections on a $\sigma$-model manifold (see equation
(\ref{genconn2}) below).

On the worldsheet, the extended supersymmetries in the left moving
sector and the right moving sector are interchanged by the worldsheet
parity transformation, $\sigma \rightarrow -\sigma$. Under this
transformation, $B \rightarrow -B$ and $S$ and $R$ are interchanged.
The matrix $Q_-$, therefore, goes over to a new matrix $Q_+$, given by
\be
\label{Qp}
\begin{array}{l}
Q_+={\ds\frac{1}{2}}\bigg(\, S+R-(S-R)\, K^T\, \bigg) \\ \\
Q_+^{-1}={\ds\frac{1}{2}}\bigg(\, S^T+R^T-(S^T-R^T)\,{\wt K}^T \bigg)
\end{array}
\ee
Since $G$ is invariant under the worldsheet parity, one expects $\wt G$
also to be invariant. In fact, using (\ref{Qm}),(\ref{Qp}) and
(\ref{SR}), it can be shown that
\be
\label{QGP}
\wt G^{-1} = Q_-\,G^{-1}\,Q_-^T=Q_+\,G^{-1}\,Q_+^T
\ee
Now, we introduce a second connection corresponding to transformations
of the form $\wt V = Q_+ V$. This connection, the relevance of which
will become clear in the next section, is given by
\be
\label{connecp}
\omega^+_\mu= G^{-1}\,{\omega^-}^T_\mu\, G ,
\ee
and, under an \odd\, deformation, transforms as
\be
\label{trconnecp}
\wt \omega^+_\mu =  Q_+\, \omega^+_\mu Q_+^{-1} -
\del_\mu Q_+ Q_+^{-1}
\ee
Having introduced two connections, we can also define the
corresponding \odd\, covariant derivatives,
\be
\label{indD}
{\cal D}^\pm_\mu = \del_\mu + \omega^\pm_\mu .
\ee
The actual form of the covariant derivative, obviously, depends on the
particular representation chosen.  At the end, note that the
transformations of the two connections, (\ref{trconnecm}) and
(\ref{trconnecp}), can also be written as
\be
\label{conn2}
\wt \omega^\pm_\mu= Q_\mp \omega^\pm_\mu Q_\pm^{-1} .
\ee

\section{Extended Supersymmetry on the Worldsheet and the $O(d,d)$
Deformations}

In this section, we obtain the transformations of the complex
structures under \odd\, and show that these deformations of the
background fields preserve the conditions of extended $(2,2)$ and
$(4,4)$ supersymmetries on the worldsheet. We will first consider the
$(2,2)$ and then the $(4,4)$ extended supersymmetries.  The result for
the left-right asymmetric situations follows trivially. We will also
comment on the possible restrictions on the form of perturbative
corrections to the \odd\, transformation equations in the context of
$(4,4)$ supersymmetry.

A non-linear $\sigma$-model with local $N=1$ supersymmetry, in both the
left and right moving sectors, can have a second supersymmetry in both
sectors provided the target manifold admits two complex
structures $J^+$ and $J^-$
\cite{ZAF}\cite{GHR}\cite{SenN=2}\cite{HP}\cite{dN}. The second
supersymmetry transformations
are obtained from the first one by replacing the worldsheet
fermions, ${\psi_\pm}^M$, by ${J^\pm}^M_{~N} {\psi_\pm}^N$ in the
supersymmetry transformation equations. The linear independence of
the two supersymmetries and the invariance of the $\sigma$-model
action imposes the following restrictions on the target manifold:
\beq
\label{alcs}
& &{J^\pm}^M_{~N} {J^\pm}^N_{~K} = -\delta^M_{~K} \\ && \nonumber\\
\label{nh}
& &{N^\pm}^K_{MN}=
{J^\pm}^L_{~M} \del^{}_{\,[L}{J^\pm}^K_{~N]}
-{J^\pm}^L_{~N} \del^{}_{\,[L}{J^\pm}^K_{~M]} = 0  \\ &&\nonumber\\
\label{hs}
& &{J^\pm}^M_{~K} G_{MN} {J^\pm}^N_{~L} = G_{KL}  \\ &&\nonumber \\
\label{cc}
& &\nabla^\pm_M {J^\pm}^N_{~K}= \del_M {J^\pm}^N_{~K} +
{\Omega^\pm}^N_{ML} {J^\pm}^L_{~K}-{\Omega^\pm}^L_{MK}{J^\pm}^N_{~L}=0
\eeq
Where, $\Omega^\pm$ are non-Riemannian connections constructed in
terms of the Christoffel symbol and the torsion tensor,
\beq
\label{genconn}
{\Omega^\pm}^N_{ML}\,&=& {\sl\Gamma}^N_{ML} \pm G^{NP}\,H_{PML} \nonumber \\
&=& \frac{1}{2} G^{NP} \left[ \del_M (G\mp B)_{PL} +
\del_L (G\mp B)_{MP} - \del_P (G\mp B)_{ML} \right]
\eeq

Equations (\ref{alcs}) - (\ref{hs}) mean that the manifold admits two
integrable almost complex structures both of which give rise to
hermitian structures. Equation (\ref{cc}) is the condition for
covariant constancy of the complex structures with respect to the
generalized connections given in (\ref{genconn}). These are the most
general conditions for the existence of (2,2) extended supersymmetry
on the worldsheet \cite{dN}. Note that in the absence of torsion, the
manifold is K\"{a}hler and the two complex structures can be chosen to
be the same. These theories have a manifest supersymmetric
representation in terms of chiral superfields \cite{ZAF}. A more
general situation arises when the torsion is not zero, but the complex
structures, $J^+$ and $J^-$, commute with each other. These theories
can be written in manifestly supersymmetric form in terms of the
chiral and twisted chiral superfields \cite{GHR}\cite{HP}. When the
two complex structures do not commute, a superfield representation is
not known in general.

To investigate the invariance of the above conditions under \odd\,
deformations, we need the transformation properties of the complex
structures $J^\pm$ under the deformations. These can be obtained by
using the transformation of the metric as given in (\ref{QGP}) and
demanding that the deformations preserve the hermitian structures
(\ref{hs}). At this level, however, there is an ambiguity in
determining the transformations of the complex structures because of
the two possible ways in which a transformation of the metric can be
written in (\ref{QGP}). One way of resolving the ambiguity, obviously,
is to find out which form of the transformations, if any, keeps the
conditions (\ref{cc}) invariant. But, a simpler and more illuminating
method is to note that for infinitesimal background fields, $G= 1+h,
B=b$, equation (\ref{eGK}) reduces to $\wt h +\wt b = R ( h+b ) S^T$.
The quantity $h+b$ can also be interpreted as the polarization tensor
in a vertex operator for the emission of massless states in
superstring theory formulated in flat backgrounds
\cite{Sen}\cite{twist}.  If we consider correlation functions of this
vertex operator which carry zero momentum along $d$ of the space
coordinates, then, the above transformation is equivalent to rotating,
by different amounts, the left-moving and right-moving parts of the
$d$ bosonic coordinates along with the left-moving and right moving
parts of their fermionic partners: $(\,\del X, \psi_+\,) \rightarrow
(\,R\del X, R\psi_+\,) $\, and $(\,\bar\del X, \psi_-\,) \rightarrow
(\,S\bar\del X, S\psi_-\,)$. In order to keep the second supersymmetry
intact, this implies that to zeroth order in the backgrounds, the
complex structures should transform as $\wt J^+=R J^+ R^T$ and $\wt
J^-=S J^- S^T$. Also in flat backgrounds, $Q_-=S$ and $Q_+=R$.
Combining this fact with (\ref{QGP}), and requiring the invariance of
the hermiticity condition (\ref{hs}), the transformations of the
complex structures under \odd\, are uniquely determined as
\be
\label{QJQ}
\begin{array}{l}
\wt J^- = Q_- J^- Q_-^{-1} \\
\wt J^+ = Q_+ J^+ Q_+^{-1}
\end{array}
\ee
These transformations trivially preserve the almost complex structures
on the manifold (\ref{alcs}). In the following, we analyze the
covariance of conditions (\ref{nh}) and (\ref{cc}). We restrict
ourselves to $J^\pm$ that are independent of the $d$ coordinates
$X^m$. For the case when this is not true, see the note added at the
end.

To simplify the analysis, we introduce two rank $3$ \odd\, tensors,
${{\cal J}^\pm}^\lambda$, defined as
\be
\label{tensors}
{{\cal J}^\pm}^{\lambda LN}_K= G^{LN} {J^\pm}^\lambda_{~K} -
{J^\pm}^L_{~K}G^{N\lambda}  -
\delta^\lambda_{~K} ({J^\pm}G^{-1})^{NL} +
\delta^L_{~K}({J^\pm} G^{-1})^{N\lambda}  ,
\ee
Under an \odd\,, they transform to
\be
\label{trtensors}
{\wt{\cal J}^\pm}\,\!^{\lambda L'N'}_{~~K'} =
{Q_\pm}^{L'}_{~L}\, {Q_\pm}^{N'}_{~N}\, {Q_\pm^{-1}}^K_{~K'} \,
{{\cal J}^\pm}^{\lambda LN}_K ,
\ee
as can be seen using (\ref{eGK}) and (\ref{QJQ}). Also note that in
terms of the induced connections $\omega_\mu^\pm$ given in
(\ref{connecm}) and (\ref{connecp}), the generalized connections
(\ref{genconn}) take the form
\be
\label{genconn2}
{\Omega^\pm}^N_{ML}={\omega^\pm_M}^N_{~L}
+(G\,\omega^\pm_LG^{-1})_M^{~N} - G^{N\mu} (G\,\omega^\pm_\mu )_{ML}
\ee
Using the above relations and the  \odd\, tensors (\ref{tensors}),
the conditions of covariant constancy of the complex structures
(\ref{cc}) can be rewritten as the following two equations (for
$M=m,\mu$),
\be
\label{newcc}
\begin{array}{ll}
M=m :\qquad &
(\nabla_m^\pm {J^\pm})^N_{~K}= (G\,\omega^\pm_\lambda)_{mL}\,
{{\cal J}^\pm}^{\lambda LN}_K =0  \\ \\
M=\mu :\qquad &
(\nabla_\mu^\pm {J^\pm})^N_{~K}=({\cal D}_\mu^\pm J^\pm)^N_{~K}+
(G\,\omega^\pm_\lambda)_{\mu L}\,
{{\cal J}^\pm}^{\lambda LN}_K =0  .
\end{array}
\ee
Here,
$$
{\cal D}_\mu^\pm J^\pm = \del_\mu J^\pm +
[\,\omega_\mu^\pm\,,\,J^\pm\,]
$$
are the two \odd\, induced covariant derivatives introduced in the
previous section (\ref{indD}). The analysis of the the invariance of
these conditions under an \odd\, deformation is now straightforward.
Using the transformation laws of the connections $\omega_\mu^\pm$
(\ref{trconnecm})(\ref{trconnecp}) and of tensors
${{\cal J}^\pm}^\lambda$ (\ref{trtensors}), along with the fact that
$(Q_\pm)^\mu_{~m}= (Q_\pm^{-1})^\mu_{~m}=0$ and $(Q_\pm)^\mu_{~\nu}=
\delta^\mu_\nu$, we obtain
\be
\begin{array}{ll}
M'=m':\qquad &
({\wt\nabla}_{m'}^\pm {{\wt J}^\pm})^{N'}_{~K'}=
{Q_\mp^{-1}}^m_{~m'}\,{Q_\pm}^{N'}_{~N}\, {Q_\pm^{-1}}^K_{~K'}
(\nabla_m^\pm {J^\pm})^N_{~K}\, =0 \\ \\
M'=\mu :\qquad &
(\wt\nabla_\mu^\pm {{\wt J}^\pm})^{N'}_{~K'}=
{Q_\pm}^{N'}_{~N}\, {Q_\pm^{-1}}^K_{~K'}
\left[\,(\nabla_\mu^\pm {J^\pm})^N_{~K}+
{Q_\mp^{-1}}^m_{~\mu} (\nabla_m^\pm {J^\pm})^N_{~K}\,\right]\,=0
\end{array}
\ee
where, the vanishing of the right-hand sides follows from
(\ref{newcc}). This proves that the transformed complex structures
are still covariantly constant with respect to the deformed
generalized connections. The last equations to check are the
integrability conditions (\ref{nh}) of the almost complex structures
$J^\pm$. After some manipulations, the \odd\, deformed Nijenhuis
tensors can be written as
$$
{{\wt N}^\pm}\,\!^{K'}_{M'N'}=
{Q_\pm^{-1}}^M_{~M'}\,{Q_\pm^{-1}}^N_{~N'}
\, {Q_\pm}^{K'}_{~K}\,
\left( {N^\pm}^K_{MN} + {{\cal J}^\pm}^{\lambda LP}_{~M}\,G_{PN}
\, \left[\, Q_\pm^{-1}\del_\lambda Q_\pm\,,\,J\,\right]^K_{~L} \right)
$$
which, on further manipulation, reduce to
\beq
\label{nhdef}
{{\wt
N}^\pm}\,\!^{K'}_{M'N'}&=&{Q_\pm^{-1}}^M_{~M'}\,{Q_\pm^{-1}}^N_{~N'}
\, {Q_\pm}^{K'}_{~K}\, \bigg(\,
{N^\pm}^K_{MN}  \nonumber\\
&&\mp\,\left(\delta^K_{~L}(GJ^\pm)_{PN}- {J^\pm}^K_{~L}G_{PN}\right)\,
\left(Q_\pm^{-1}\,(S-R)\right)^{Lm}\, (\nabla_m^\pm J^\pm)^P_{~M}
\,\bigg) =0
\eeq
Here, the vanishing of the deformed Nijenhuis tensors follows from
equations (\ref{nh}) and
(\ref{cc}). Thus, the deformed almost complex structures are
integrable. This completes the proof of the invariance of the $(2,2)$
extended supersymmetry on the worldsheet under \odd\, deformations.

The only other possible extension of the $N=1$ supersymmetry is to
$N=4$ \cite{GHR}\cite{HP}. In the $(4,4)$ case, this extension
requires the existence, in each chiral sector separately, of three
complex structures, $J_a; a=1,2,3$. Each one of the $J_a$'s satisfies
conditions (\ref{alcs})-(\ref{cc}) independently.  In addition, the
$J_a$ satisfy an $SU(2)$ algebra, giving rise to a quaternionic
structure on the target manifold,
$$
J^\pm_a J^\pm_b=-\delta_{ab} + \epsilon_{abc} J^\pm_c .
$$
Also, condition (\ref{nh}) is now generalized to the vanishing of the
mixed Nijenhuis tensors,
$$
N^{\pm\,K}_{(ab)\,MN}= J^{\pm\,L}_{(a\,M}\,
\del^{}_{[L}\,J^{\pm\,K}_{b)\,N]} -J^{\pm\,L}_{(a\,N }\,
\del^{}_{[L}\,J^{\pm K}_{b)\,M]} = 0
$$
The invariance of the
constraints of extended $(4,4)$ supersymmetry under a deformation
follows from the above discussion for the $(2,2)$ case coupled with
the fact that the transformations of the complex structures do not
affect the $SU(2)$ index $a$. This proves the invariance of the
extended $(4,4)$ supersymmetry under \odd\, deformations when all
complex structures are independent of the $d$ coordinates $X^m$. Since
supersymmetry is preserved in the left and right chiral sectors
independently, the above analysis can be trivially generalized to any
model with extended $(p,q),\, p,q=0,1,2,4\, $ supersymmetry.

The invariance of the $(4,4)$ supersymmetry under the deformations
may have an implication for the possible form of higher $\sigma$-model
loop corrections to the \odd\, transformations. The form of the
transformation given in (\ref{oddM}), with $M$ defined as in (\ref{M}),
is correct to one-loop in the $\sigma$-model perturbation theory.
However, the arguments in \cite{Sen},\cite{twist} and the analysis of
\cite{GR} and \cite{Def} imply the existence of corrections, to all
orders, to the transformation. On the other hand, if the
non-renormalization theorems for the extended $(4,4)$ supersymmetry on
the worldsheet \cite{HP} are valid, then, for these theories, the
corrections to the \odd\, transformations must vanish. This restricts
the possible form of the higher loop corrections in terms of the
constraints which the $(4,4)$ supersymmetry imposes on the background
fields.

\section{Complex Structures, Duality and Manifest $N=2$ Supersymmetry}

In this section we analyse the transformations of the complex
structures and show that, starting from a theory with manifest $N=2$
supersymmetry, the deformed theories do not, in general, admit a
manifestly supersymmetric description in terms of chiral and twisted
chiral superfields. We write the action of a general discrete duality
transformation on the complex structures and discuss the case of
K\"{a}hler manifolds.

The existence of extended $N=2$ supersymmetry on the worldsheet
severely constrains the target space geometry by requiring the
existence of two complex structures $J^\pm$ satisfying equations
(\ref{alcs})-(\ref{cc}). A simpler situation arises when the torsion,
$H_{MNP}$, on the target space is zero. Conditions
(\ref{alcs})-(\ref{cc}), then, do not distinguish between the two
complex structures which can, therefore, be chosen to be the
same. This complex structure is covariantly constant with respect to
the Riemmanian connection and the corresponding manifold is
K\"{a}hler.  A non-linear $\sigma$-model defined on a K\"{a}hler
manifold can be written, in a manifestly $(2,2)$ supersymmetric form,
in terms of the K\"{a}hler potential as a function of $N=2$ chiral
superfields. In the presence of torsion, a superfield formulation of
the theory is not always known. However, there is a special class of
theories with $H_{MNP}\neq 0$ which still admit a manifestly
supersymmetric description in terms of chiral and twisted chiral
superfields. These are the theories in which the two complex
structures commute and are, therefore, simultaneously
diagonalizable \cite{GHR}\cite{HP}. We can always choose a canonical
basis in which $J^-$
takes the form ${J^-}^a_{~b}=i \delta^a_{~b}\,,\, {J^-}^{\bar
a}_{~\bar b}= -i \delta^{\bar a}_{~\bar b}$, where, $a, \bar a =1,
\dots , D/2$ label the holomorphic and antiholomorphic coordinates.
In this basis, $J^+$ is also diagonal but the arrangement of $\pm
i$\,'s on the diagonal depend on the number of twisted chiral
superfields or, equivalently, the form of $B_{MN}$. Such a manifold is
said to have a product structure.

The K\"{a}hler and product structures, however, are very special and
are not, in general, preserved under non-trivial \odd\, deformations
of the complex structures. To investigate this point, we first
consider the case of flat backgrounds. In this case, as discussed
above equation (\ref{QJQ}), the transformations can be interpreted as
independent $d$-dimensional rotations of the left and right moving
parts of the bosonic and fermionic coordinates inside correlation
functions which carry zero momentum along $d$ of the space directions.
Modulo ordinary $d$-dimensional rotations and parity transformations,
which are symmetries of the theory, we can choose $S =1$ in
(\ref{QJQ}).  The complex structures then transform as $J^-\rightarrow
J^-\,, J^+\rightarrow RJ^+R^T$. Note that $R$ is rotation a involving
the real coordiantes. Therefore, if we choose the usual real
representation for $R$, then, $J^+$ also has to be written in the real
coordinate basis. If we write $J^+$ in the diagonal basis (which is
not real), then $R$ also has to be transformed appropriately.  It can
be explicitly checked that starting from commuting complex structures,
non-trivial $R$ transformations that preserve this commuting nature
are ({\it i}) the ones which flip the $+i$ and $-i$ eigenvalues of
$J^+$ written in the canonical basis of $J^-$, ({\it ii}) the ones
that correspond to $O(2)$ rotations involving the real and imaginary
parts of the same complex coordinate (note that a rotation which mixes
the real and imaginary parts of different complex coordinates does not
qualify). The first case corresponds to converting a chiral superfield
into a twisted chiral one, or {\it vice versa} and it can be checked
that these transformations form the discrete duality subgroup
(\ref{dual}) of $O(d,d,R)$. This once again shows the connection
between the usual duality and the $N=2$ duality of \cite{GHR}. Case
({\it ii}) corresponds to $O(2)\times O(2)/O(2)$ deformations.  The
remaining elements of $O(d)\times O(d)/O(d)$ do not, in general,
preserve the commuting nature of the complex structures even in flat
background fields. Therefore, in the following, where the background
fields are not flat, we will concentrate only on cases ({\it i}) and
({\it ii}) above.

In the presence of non-trivial background fields, it can be explicitly
checked that $O(2)\times O(2)/O(2)$ deformations do not generically
preserve the commuting nature of the complex structures. One exception
is when the manifold is K\"{a}hler and the metric is independent of
one of the complex coordinates, say $z_1$, and its conjugate. then it
can be explicitly checked that an $O(2,2)$ deformation involving only
the real and imaginary parts of $z_1$ does not change the complex
structures. The manifold, thus, remains K\"{a}hler even after the
deformation.  We, now, turn to the case of discrete duality
transformations (\ref{dual}). A duality with respect to $d$ of the
coordinates, on which the background fields do not depend, is
generated by ${\cal S}=1\,, {\cal R}=1-2\, \sum_{i=1}^d \varepsilon_i$
\cite{Busc}\cite{RV}\cite{GR}\cite{GRVM}. The dual complex structures,
obtained from (\ref{QJQ}), are given by
\be
\label{dualJ}
{\wt J}^-=\left( \begin{array}{cc}
(KJ^-)^m_{~l} ({\cal K}^{-1})^l_{~n} &
-(KJ^-)^m_{~l} ({\cal K}^{-1})^l_{~p} K^p_{~\nu}+ (KJ^-)^m_{~\nu}\\ & \\
{J^-}^\mu_{~l}({\cal K}^{-1})^l_{~n}&
-{J^-}^\mu_{~l}({\cal K}^{-1})^l_{~p} K^p_{~\nu}+{J^-}^\mu_\nu
\end{array}\right)
\ee
and
\be
{\wt J}^+=\left( \begin{array}{cc}
(K^TJ^+)^m_{~l} ({\cal K}^{-1 T})^l_{~n}&
(K^TJ^+)^m_{~l} ({\cal K}^{-1T})^l_{~p} {K^T}^p_{~\nu}
-(K^TJ^+)^m_{~\nu}\\ & \\
-{J^+}^\mu_{~l}({\cal K}^{-1T})^l_{~n}&
-{J^+}^\mu_{~l}({\cal K}^{-1T})^l_{~p} {K^T}^p_{~\nu}
+{J^+}^\mu_\nu
\end{array}\right)
\ee
where, ${\cal K}_{mn}=K_{mn}\,;m,n=1,\dots,d$ is a $d\times d$
submatrix of $K$, and the indices of $K$ are raised and lowered by a
flat metric which appears in the \odd\, transformations and has not
been explicitly written here. For $d=1$, these expressions were
obtained in \cite{KR}.  From the above, one can see that even if the
complex structures, $J^+$ and $J^-$, commute, their duals in general
do not commute except for some restricted backgrounds. A situation
which can be discussed in some generality is the K\"{a}hler manifold
with isometries, on which, the two complex structures are chosen to be
equal. If the isometries are along some of the real coordinates which
define the canonical basis for the complex
structure, then, one can explicitly check that the dual complex
structures are still commuting. This reproduces the duality of $N=2$
theories discovered in \cite{GHR} when the starting theory is defined
on a K\"{a}hler manifold. However, for more general isometries, after
a duality transformation with respect to $d$ isometries, the dual
complex structures do not commute, even for this restricted class. The
non-zero terms in the commutator are, however, proportional to the
components of the K\"{a}hler form, $GJ$, in the isometry
directions. Since the K\"{a}hler form is antisymmetric, for $d=1$ the
dual complex structures commute without further restrictions on the
background fields.

{}From the above discussion it follows that though the deformations
preserve the extended $N=2$ supersymmetry, the K\"{a}hler and product
structures are not, in general, preserved. In such cases, the deformed
theories do not have a formulation in terms of chiral and twisted
chiral superfields. This implies that, starting from a conformally
invariant non-linear $\sigma$-model with manifest extended $N=2$
supersymmetry and some isometries, one can construct a large class of
new theories with $N=2$ extended supersymmetry, for which, a
superfield representation may not be known. This is achieved without
solving constraints (\ref{alcs})-(\ref{cc}) and the $\beta$-function
equations. Duality transformations applied to the models of \cite{GHR}
preserve the superfield representation. The discussion also clarifies
the connection between the duality transformations (\ref{dual}) and
the duality of $N=2$ theories discussed in \cite{GHR}.

\section{The $O(d,d+16)$ Deformations and Extended Worldsheet
Supersymmetry in Heterotic String Theory}

In this section we rewrite the action of the \oddh\, group on the
heterotic string theory backgrounds, including the gauge
Chern-Simons term, in terms of the metric vielbein. We then construct
a connection induced by these transformations and show that the
extended worldsheet supersymmetry in heterotic string theory is
preserved under the deformations.

In heterotic string theory, the \odd\, group of deformations of the
string vacua with $d$ isometries is enlarged to an $O(d,d+p)$ group
provided all the background gauge fields belong to a subgroup that
commutes with $p$ of the Cartan generators of the gauge group
\cite{Narain}\cite{twist}\cite{JMJS}. For simplicity, we consider
the case when $p=16$ and, therefore, all background gauge fields are
abelian. As
in the non-heterotic case, the action of the \oddh\, group on the
background fields is given in terms of a $(2D+16)\times(2D+16)$
dimensional matrix $M$ constructed from $G_{MN}\,,\, B_{MN}$ and the
gauge fields $A^I_M$, where, $I=1,\dots,16$ is the gauge group
index \cite{twist}. The matrix $M$ transforms under the adjoint action
of the
\oddh\, group embedded in the fundamental representation of
$O(D,D+16)$. The transformations of the background fields are uniquely
determined from that of $M$. At the one-loop level, this is also
accompanied by a transformation of the dilaton field $\Phi$. As in
section $2$, we rewrite the transformations, in a slightly modified
form, in terms a $(2D+16)\times (D)$-dimensional matrix $\xi$ defined
as
\be
\xi =
\left(\begin{array}{c}
e \\ K\,e \\ - A\,e
\end{array} \right)
\ee
where, $A$ is a $16\times D$-dimensional matrix and $K$ is given by
\be
K_{MN}=G_{MN}+B_{MN}+{1\over 2}\,A_M^I\,A_N^I
\ee
The matrix $M$, mentioned above, can now be constructed in terms of
$\xi$ as\footnote{The fields here are related to those of
\cite{twist} by $B\rightarrow -B, K\rightarrow -K, A\rightarrow
-A/{\sqrt 2}$.}
\be
M = \xi\, \eta\, \xi^T =
\left(\begin{array}{c} e \\ Ke \\ -Ae \end{array}\right)\!\!
\begin{array}{c}\eta\\ \, \\ \, \end{array}\!\!
\begin{array}{c}
\left(\begin{array}{ccc} e^T & e^T K^T &-e^T A^T \end{array}\right)
\\ \, \\ \,
\end{array}
\ee
Under an \oddh\, transformation, $\xi$ transforms as
\be
\label{oddh}
\wt\xi =\Omega \xi\,,\quad \Omega \in O(d,d+16) \subset O(D,D+16)
\ee
where, the representation is chosen such that the defining equation
for $\Omega$ takes the form
\be
\label{Lhet}
\Omega^T\, L\,\Omega = L \,,\quad
L= \left(\begin{array}{ccc}0&1_D&0\\ 1_D&0&0\\0&0&-1_{16}\end{array}
\right)
\ee
These transformations include the $GL(d,R)$ transformations and gauge
transformations of $B_{MN}$ and $A^I_M$. The non-trivial deformations
are generated by the elements from $(O(d+16)$$\times$$O(d))/O(d)$. These
are parametrized as
\be
\Omega = {1\over 2}\left(\begin{array}{ccc}
S+R & S-R & -R_1^T \\
S-R & S+R & R_1^T  \\
-R_2& R_2 & 2R_3 \end{array}\right)
\ee
modulo the subgroup generated by $S=R$. Here, $S$ and $R$ are again
given by equation (\ref{SR}) although it is no longer necessary to have
$R\in O(d)$; $R_3$ is a
$16\times 16$-dimensional matrix and $R_{1,2}$ are
$16\times D$-dimensional matrices of the form
$$
R_{1,2}=\left(\begin{array}{cc}
\sqrt{2}\,{\cal R}_{1,2}& 0_{16\times (D-d)} \end{array}\right)
\,,\quad\hbox {where},
\left(\begin{array}{cc}{\cal R}&{\cal R}^T_1 \\
{\cal R}_2 & R_3 \end{array}\right) \in \,O(d+16)
$$
Under the deformations generated by the above elements, the background
fields transform as
\be
\label{eGKA}
\begin{array}{cl}
{\wt e} & = Q\,e  \\
{\wt K}& ={\ds{1\over 2}}\bigg( (S-R) + (S+R)\,K - R_1^T\, A \bigg)\,
Q^{-1}\\
{\wt A} & ={\ds{1\over 2}}\bigg( R_2 - R_2\,K + 2\, R_3\, A \bigg)\,
Q^{-1}
\end{array}
\ee
where, $Q$ is given by
\be
\label{Qhet}
Q= {1\over 2}\bigg( (S+R) + (S-R)\,K +R_1^T\,A \bigg)
\ee
The transformation of the metric can be obtained from that of the
inverse vielbein $e$.

Now, as in section $2$, we introduce an \oddh\, induced connection as
follows. Note that, under \oddh\, deformations (\ref{oddh}), the
quantity $\xi^T\,L\,\del_\mu\,\xi$ is invariant and, from
(\ref{eGKA}), the quantity $e\,\del_\mu\,e^{-1}$ transforms as a
connection corresponding to transformations of the type $\wt V = Q
V$. Using these two quantities, we construct a non-minimal \oddh\,
induced connection $\omega_\mu$ as
\be
\omega_\mu = \frac{1}{2}\, e\eta\xi^T L \del_\mu\xi e^{-1}
- e\del_\mu e^{-1} ,
\ee
which, in a more recognizable form, can be written as
\be
\label{connhet}
\omega_\mu ={1\over 2}\,G^{-1}\,\big( \del_\mu\,K - A^T\,\del_\mu\,A
\big)
\ee
Under an \oddh\, deformation, $\omega_\mu$ transforms to
\be
\wt \omega_\mu =  Q\, \omega_\mu Q^{-1}- \del_\mu Q\, Q^{-1}
\ee
Using the above construction, we set out to prove the invariance of the
extended worldsheet supersymmetry in heterotic string theory under the
\oddh\, deformations.

In heterotic string theory, the local $(0,1)$ worldsheet supersymmetry
can be extended to a $(0,2)$ supersymmetry in a way similar to the
superstring case \cite{HW}\cite{SenN=2}\cite{HP}. The target
manifold is required to admit an almost
complex structure $J$ with vanishing Nijenhuis tensor, $N^K_{MN}=0$,
and a hermitian metric, $J^T\,G\,J=G$. The difference with the
superstring case, however, is that the torsion tensor $H_{MNK}$ now
also contains a gauge Chern-Simons term\footnote{In general, the
torsion also contains the Lorentz Chern-Simons term which is a higher
derivative term and will not be considered here.} coming from the
one-loop chiral anomaly, $H_{MNK}=(1/2)(\del_M B_{NK}+\cdots)+(1/4)(
A^I_M F^I_{NK}+\cdots)$, where, the dots denote cyclic permutations
and we have restricted ourselves to abelian gauge fields. This
modifies the generalized connection and, thus, the condition of
covariant constancy of the complex structure. With this modification
in mind, the conditions for the existence of extended supersymmetry in
heterotic string theory are again given by equations
(\ref{alcs})-(\ref{cc}) for $J=J^-$. The only extra condition is the
constraint on the gauge field background,
\be
\label{fj}
F^I_{KL}\,J^L_{~M}-F^I_{ML}\,J^L_{~K}=0
\ee
The transformation of the complex structure $J$ under an \oddh\,
deformation is obtained by requiring the covariance of the hermiticity
condition of the metric and is given by $\wt J=Q\,J\,Q^{-1}$, with $Q$
as given by (\ref{Qhet}). This also preserves $J^2 =-1$. To
investigate the covariance of the remaining conditions, it is
convenient to write them in terms of the induced connection
$\omega_\mu$\, (\ref{connhet}) and the rank $3$ tensor, ${\cal
J}^{\lambda LN}_K ={{\cal J}^-}^{\lambda LN}_K$\,
(\ref{tensors}). First, consider the covariant constancy condition
$\nabla_M J=0$. It turns out that, even in the presence of the gauge
Chern-Simons term, the relation between the generalized connection
$\Omega^-$ and the \oddh\, induced connection (\ref{connhet}) is given
by equation (\ref{genconn2}). Therefore, the covariant constancy
condition again takes the form given in (\ref{newcc}). Next, we
consider the condition on the gauge fields (\ref{fj}). It is easily
seen that this equation can be rewritten as
\be
\label{newfj}
\del_\lambda\,A^I_{~L}\,{\cal J}^{\lambda LN}_K\,G_{NM}=0
\ee
Using the above equations in terms of the \oddh\, covariant variables,
a straightforward calculation shows that, after a deformation, the
tensors $N^K_{MN}, \nabla_M J$ and the left hand side of equation
(\ref{newfj}) transform into linear combinations of each other and,
therefore, remain equal to zero. This proves the invariance of the
$(0,2)$ worldsheet extended supersymmetry in heterotic string theory
under the \oddh\, deformations of the background fields. The
generalization to $(0,4)$ supersymmetry follows from the discussion,
in section $3$, of extended $(4,4)$ supersymmetry in superstring
theory.

\section{Conclusions}
We have shown that the $O(d,d;{\rm R})$ deformations of the
superstring vacua and the $O(d,d+16;{\rm R})$ deformations of the
heterotic string vacua can be rewritten, in a simpler way, in terms of
the target space vielbeins. Though, these transformations are global
in the space-time sense, their non-linear action on the background
fields enables us to construct some induced space-time connections. We
write down the transformations of the complex structures associated
with the extended worldsheet supersymmetries, under the above
deformations. The analysis is valid only when the complex structures
are independent of the $d$ coordinates with respect which the
deformations are performed.  Using the induced connections and some
tensors which transform covariantly under the deformations, we show
that the $O(d,d;{\rm R})$ deformations of the superstring vacua and
the $O(d,d+16;{\rm R})$ deformations of the heterotic string vacua
preserve the extended supersymmetries on the worldsheet. They, therefore,
generate marginal deformations of the associated superconformal field
theories. In the case of extended $(2,2)$ supersymmetry, we discuss
the transformations of the complex structures in relation to the
superfield representation of the deformed theories. It is shown that
generic deformations do not preserve the known superfield
representations of the theories in terms of chiral and twisted chiral
superfields. We also write down, explicitly, the transformations of
the complex structures under the target space duality. The discussion
clarifies the relation between the above duality transformation and a
duality in the $N=2$ theories discussed by Gates, Hull and Ro\v{c}ek
\cite{GHR}. We also comment on the possible form of the perturbative
corrections to the \odd\, transformations in the context of $(4,4)$
theories.

\noindent{\bf Acknowledgement}

It is a pleasure to thank Ashoke Sen and Suresh Govindarajan for
discussions and comments.

\centerline{\bf Note Added}
When a complex structure depends on the coordinates with respect to
which an \odd\, transformation is performed, then the extended
supersymmetry in the deformed theory is no longer realized in the standard
way. This issue was addressed in \cite{BS,H} for the case of T-duality
trasformations and it was shown that in the dual theory, the extended
supersymmetry is realized non-locally. The generalization of this
discussion to \odd\, transformations is straightforward.

\end{document}